# Millimeter wave to terahertz compact and low-loss superconducting plasmonic waveguides for cryogenic integrated nano-photonics

Samane Kalhor [1,2], Majid Ghanaatshoar [1], Hannah J. Joyce [3], David A. Ritchie [4], Kazuo Kadowaki [5], and Kaveh Delfanazari [2,3,4]*

[1] *Laser and Plasma Research Institute, Shahid Beheshti University, G.C., Evin, 1983969411, Tehran, Iran*

[2] *James Watt School of Engineering, University of Glasgow, Glasgow G12 8QQ, UK*

[3] *Electrical Engineering Division, University of Cambridge, Cambridge CB3 0FA, UK*

[4] *Department of Physics, Cavendish Laboratory, University of Cambridge, Cambridge CB3 0FA, UK*

[5] *Division of Materials Science, Faculty of Pure & Applied Sciences, University of Tsukuba 1-1-1, Tennodai, Tsukuba, Ibaraki 305-8573, Japan*

\* Corresponding author: kaveh.delfanazari@glasgow.ac.uk

**Abstract**

Plasmonic, as a rapidly growing research field, provides new pathways to guide and modulate highly confined light in the microwave to the optical range of frequencies. We demonstrate a plasmonic slot waveguide, at the nanometer scale, based on high transition temperature ($T_c$) superconductor $Bi_2Sr_2CaCu_2O_{8+\delta}$ (BSCCO), to facilitates the manifestation of the chip-scale millimeter waves (mm-waves) to terahertz (THz) integrated circuitry operating at cryogenic temperatures. We investigate the effect of geometrical parameters on the modal characteristics of the BSCCO plasmonic slot waveguide between 100 GHz and 500 GHz. In addition, we investigate the thermal sensing of the modal characteristics of the nanoscale superconducting slot waveguide and show that at a lower frequency, the fundamental mode of the waveguide has a larger propagation length, a lower effective refractive index, and a strongly localized modal energy. Moreover, we find that our device offers a larger SPP propagation length and higher field confinement than the gold plasmonic waveguides at broad temperature ranges below BSCCO's $T_c$. The proposed device can open up a new route towards realizing cryogenic low-loss photonic integrated circuitry at the nanoscale.

**Introduction**

The high transition temperature ($T_c$) superconducting $Bi_2Sr_2CaCu_2O_{8+\delta}$ (BSCCO) intrinsic Josephson junctions (IJJs) based THz emitters radiate intense, coherent, and continuous THz photons with frequencies ranging from 0.1 to 11 THz [1-27]. Such THz devices can also be used as surface current sensitive detectors due to the unique electrodynamics of BSCCO quantum material. Therefore, BSCCO-based devices are valuable for many applications, including THz imaging, interferometry, and absorption measurement [28]. The design of low-loss mm-wave to THz components, e.g., waveguides, capable of being integrated with such superconducting emitters, and detectors are vital to accomplishing all BSCCO-made chip-integrated mm-waves to THz circuitry. Moreover, the exploitation of superconducting quantum materials into the architectures of waveguides enables the implementation of real-time sensing and controlling of waveguides, due to the sensitivity of quantum mechanical phases of superconductors to external stimuli such as magnetic field, temperature, light, and current [29]. Cooper pairs in superconductors have an equivalents response to that of electrons in plasmonic metals at high frequencies.

Plasmonics deals with propagating surface plasmon polariton (SPPs) as the coupled oscillation of electrons and electromagnetic waves [30]. The innovative physical effect of plasmonic devices such as subwavelength localization of the electromagnetic field opens up a new route in novel chip-scale integrated photonic devices [31]. It was shown that $YBa_2Cu_3O_7$ (YBCO) and niobium (Nb) plasmonic superconducting waveguides offer the superior long plasmon propagation distance in comparison to noble metals at THz frequencies [32-33] due to the intrinsic low-loss plasmonic properties of superconductors.

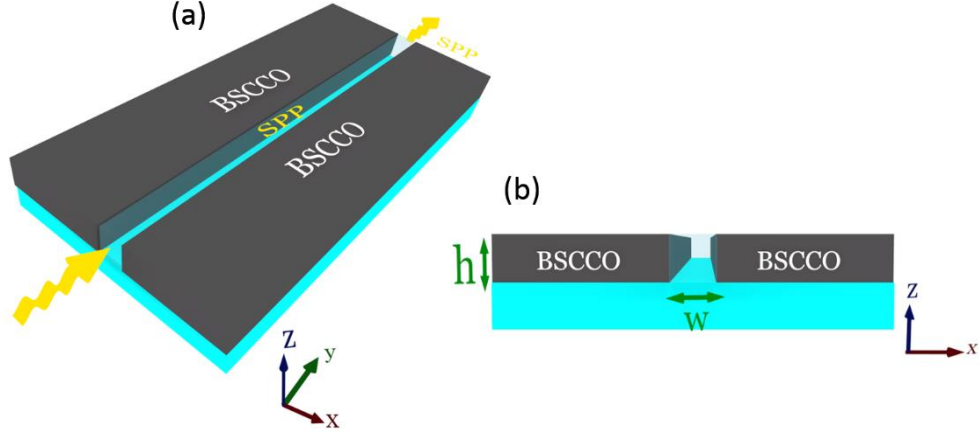

**Figure 1**: (a) 3D Schematic diagram of the proposed BSCCO based plasmonic slot waveguide, (b) the cross-sectional view of the waveguide at x-z plane.

In this paper, we propose mm-waves to THz superconducting plasmonic slot waveguide (PSW) based on BSCCO. We first study the mode characteristics of the BSCCO PSW, including the effective refractive index, the propagation loss of SPPs, and mode energy confinement. Furthermore, we investigate thermal tuning of the modal characteristics of such waveguides at temperature ranges between $T$=10 K and 100 K at the selected frequencies of $f$=0.1 THz, 0.3 THz, and 0.5 THz. The proposed waveguide can be integrated with BSCCO based THz sources and detectors. It is also suitable for various applications such as tunable modulators and photodetectors.

**Structure Design and Methods**

The 3D schematic of the proposed BSCCO PSW is shown in Figure 1(a). The cross-sectional view of the 3D SPW at the x-z plane in Figure 1(b) shows that the waveguide consists of a deep subwavelength air slot of width $w$ in a thin film of BSCCO with a thickness of $h$. The generated SPPs propagate through the air slot.

We employ the numerical finite element simulation method to calculate the eigenmodes of the plasmonic waveguide at a specific frequency $\omega$. Here an $\exp(-i\beta y)$ dependence for electric field is considered because waveguide is uniform along the y-direction [34]. Therefore, electric field $E$ distribution in the waveguide can be written as

$$\boldsymbol{E}(x,y,z) = E(x,z)\exp(-i\beta y) \quad (1)$$

where $\beta = \beta_1 + i\beta_2$ is the complex propagation constant of the waveguide's mode.

Inserting this electric field into the electromagnetic wave equation [35]

$$\nabla \times \nabla \times \boldsymbol{E} = \frac{\omega^2}{c^2}\varepsilon\boldsymbol{E} \quad (2)$$

and solving it as an eigenvalue problem allows us to retrieve the modal characteristics of the waveguide, including the real part of the effective refractive index ($N_{eff}$) and propagation length ($L_p$) of SPPs for the fundamental mode of BSCCO PSW from equations [34][36]:

$$N_{\text{eff}} = \beta/k_0, \quad (3)$$
$$L_p = 1/(2\,\text{Imag}(\beta)), \quad (4)$$

where $k_0$ is the free space wavevector and $c$ is the speed of light in the vacuum. $N_{eff}$ is an indicator of the localization of SPP's energy and wavelength. In addition, the dispersion relation of the waveguide is defined as $\omega = \omega(\beta)$ [34].

In equation (2), $\varepsilon$ is the permittivity of the relevant medium. Permittivity of air $\varepsilon_{air}$ is 1, and the temperature and frequency-dependent a-b plane complex conductivity of BSCCO film with $T_c$= 85 K is extracted from the experimental THz time-domain spectroscopy data [37-39]. The complex permittivity of BSCCO can be obtained from its complex conductivity [39].

For obtaining the eigenmode of the waveguide, the area of computation is considered to be large enough, and also perfectly matched layer (PML) absorbing boundary conditions are used along the x and z-axis. Therefore, the reflection of fields from the boundaries is negligible.

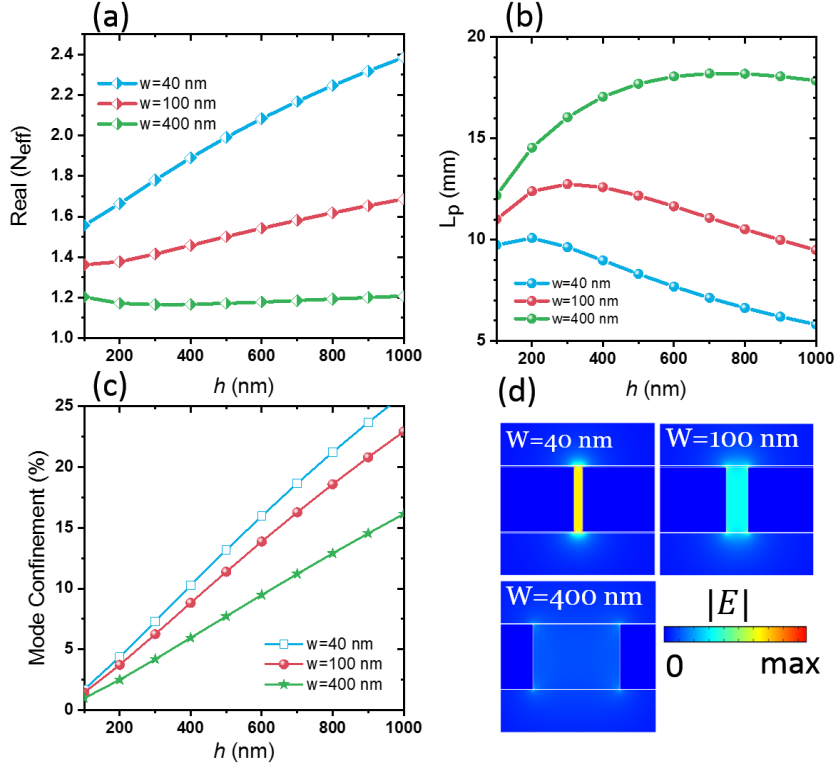

**Figure 2**: (a) Real part of the mode effective refractive index ($N_{eff}$), (b) propagation length ($L_p$), and (c) mode confinement of BSCCO THz plasmonic waveguide as a function of BSCCO thickness $h$ for three slot widths of $w$=40, 100, and 400 nm at $f$=0.1 THz, and $T$=10 K. (d) electric field distribution at slot width $w$=40, 100, and 400 nm for BSCCO height $h$=300 nm. All field distributions curves have the same color bar.

In addition, the factor of mode confinement is calculated as a ratio of power flow in the slot ($w \times h$ area) to the total power flow in the waveguide normal to the *x-z* plane [36].

$$\Gamma = \frac{\int_{slot} Re\{E \times H^* \cdot n\} \, dA}{\int_{total} Re\{E \times H^* \cdot n\} \, dA} \quad (5)$$

where $E$ and $H$ are the electric and magnetic field vectors, respectively, and $n$ is the normal unit vector in the *y*-direction.

## Results and Discussion

The highest possible mode quality of the waveguide is obtained through optimization of the slot width $w$ and BSCCO thin film thickness $h$ at the temperature $T$=10 K and frequency $f$=0.1 THz. The modal characteristics are controllable by the structural size of the waveguide. The effective refractive index ($N_{eff}$) and propagation length ($L_p$) of BSCCO PSW as a function of BSCCO height $h$ for different slot widths $w$ are shown in Figures 2(a)-(b).

At each BSCCO/air interface within the air gap, SPPs are formed. These two formed SPPs are coupled and create a transverse electromagnetic (TEM) wave mode. $N_{eff}$ is larger than the air refractive index and is adjustable by the structural size. As $h$ increases, $N_{eff}$ gets larger, but $L_p$ decreases because the superconductor/air interface height in the slot increases. The larger portion of BSCCO (whose $N_{eff}$ is higher than air) results in larger $N_{eff}$. Besides, as width $w$ gets smaller, $N_{eff}$ increases, and $L_p$ reduces. Once the slot width is narrow, the SPP related to the two BSCCO surfaces form the coupled SPPs [40]. Therefore, as the $w$ of the slot decreases, the propagation constant β increases and leads to the rising of $N_{eff}$ and the reduction of $L_p$ [34]. The fall of $L_p$ at very low $h$ arises from the decoupling of two formed SPPs.

The mode confinement of SPPs is shown in Figure 2(c). It determines the enhancement of energy in the slot region. The mode confinement reduces with increasing the slot width *w*. To clarify this, we show the electric field distribution at different slot widths *w* for a constant *h*=300 nm in Figure 2(c). We see that slot width of *w*=400 nm has the lowest electric field distribution within the slot. Even though the narrower slot dimension has large energy confinement, but it suffers from the lower propagation length. There is a tradeoff between the energy confinement of SPPs within the slot and SPP's propagation length. The largest $L_p$ for *h*=300 nm thick SPW is for slot width of *w*=100 nm. Hence, these values (*w*=100 nm and *h*=300 nm) are chosen as SPW optimum dimensions. Based on this optimization, $N_{eff}$ is 1.42. The mode has a shorter wavelength in comparison to the free space. Therefore, the SPP's wavelength, which is defined as $\lambda_0/N_{eff}$, is 2.1 mm, and the SPP's field is confined in the air slot as small as $\lambda_0^2/(3 \times 10^8)$. Here, $\lambda_0$=3 mm is the free-space wavelength.

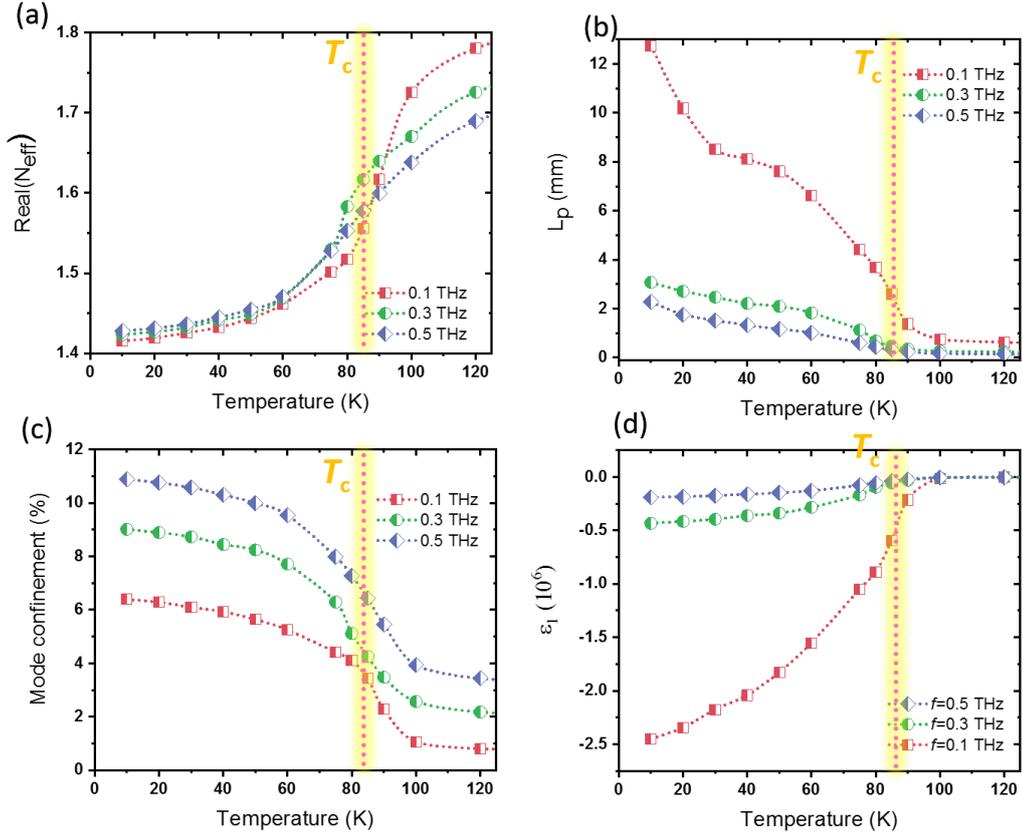

**Figure 3**: (a) Real part of the mode effective refractive index ($N_{eff}$), (b) propagation length of SPPs ($L_p$), (c) the mode confinement of the PSW as a function of temperature, and (d) the real part of BSCCO permittivity at different frequencies. The vertical dashed lines show the $T_c$ of BSCCO.

The modal characteristics of the waveguide are dependent on the plasmonic properties of BSCCO. $N_{eff}$, $L_p$, and Mode confinement are shown in Figure 3(a)-(c) as a function of temperature for three different frequencies of *f*=0.1, 0.3, and 0.5 THz. For each frequency, it is found that $N_{eff}$ reduces but $L_p$ and mode confinement increase significantly as BSCCO enters the superconducting state below $T_c$ (the vertical dashed line). The magnitude of the real part of BSCCO permittivity ($|\varepsilon_1|$) increases below $T_c$ (see Figure 3(d)). The continuity of normal component of electric field displacement (D) at the boundary of BSCCO and air interface as $\varepsilon_{BSCCO} E_{BSCCO\perp} = \varepsilon_{air} E_{air\perp}$ results in the decrease of the electric field in BSCCO, by increasing $|\varepsilon_1|$, due to the material temperature reduction. Here, $\varepsilon_{BSCCO}$ and $\varepsilon_{air}$ are permittivity of BSCCO and air, respectively. $E_{1\perp}$ and $E_{2\perp}$ are normal components of the electric field in BSCCO and air, respectively. The electric field reduction in the BSCCO results in lower modal propagation constant β and lower $N_{eff}$. The growth of $L_p$ and mode confinement with cooling the waveguide is also the outcome of lower β.

Figure 3(d) shows the real part of BSCCO permittivity at different frequencies. Indeed, superconductors are intrinsically plasmonic media with a negative real part of complex permittivity. Superconductor plasmonic properties are determined by the coexistence of normal and superconducting plasma. Normal carriers are responsible for scattering processes. At $T_c$ and above, all carriers are in the normal phase. With a reduction in temperature to below $T_c$, the ratio of superconducting carriers to normal carriers increases. At zero temperature, all carriers turn into supercarriers according to the well-known two-fluid model [41-42]. Above $T_c$ (vertical dashed line in Figure 3(d)) in the normal state of BSCCO, $\varepsilon_1$ has a very low value. The material is nevertheless in the plasmonic regime, with a huge loss. At cryogenic temperatures below $T_c$ of BSCCO, the absolute value of $\varepsilon_1$ increases. Therefore, loss of the material decreases due to the growing of supercarrier densities. The growth of SPP's propagation length by reducing the temperature in Figure 3(b) is a result of loss reduction in BSCCO below $T_c$.

From Figure 3(a)-(c), it could also be found that $N_{eff}$ increases and $L_p$ decreases as the frequency increases. $N_{eff}$ growing (Figure 3(a)) is due to the reduction of the absolute value of the real part of BSCCO permittivity with frequency (See Figure 3(d)), and also larger penetration of mode power in BSCCO. Larger modal penetration in BSCCO means lower mode confinement within the slot (see Figure 3(c)). The reduction of $L_p$ with increasing frequency in Figure 3(b) is a result of increasing the Ohmic loss and also due to the larger fraction of the modal power in BSCCO.

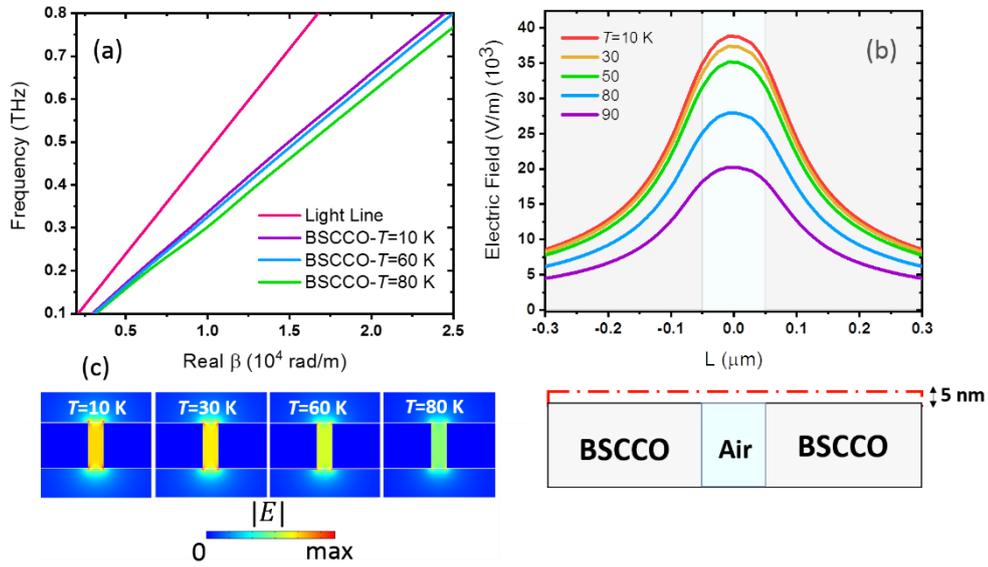

**Figure 4**: (a) Dispersion curves for the mode of the BSCCO PSW at $T$=10 K (purple), $T$=60 K (blue), and $T$=80 K (green). The red line shows dispersion relation of the THz waves propagating in free space, (b) electric field intensity at the red dashed line 5 nm above the waveguide, at different temperatures between $T$= 10 K and 90 K, at $f$=0.1 THz. Cross section of the waveguide shows the relative position of BSCCO film and the slot area in the electric field curves. The pale blue area shows the gap while the grey area shows the BSCCO part, and (c) the electric field distribution of the BSCCO THz PSW for width $w$=100 nm and height $h$=300 nm at $f$=0.1 THz for selected temperatures $T$=10 K, 60 K, and 80 K. All field distributions have the same color bar.

For clarifying the effect of temperature, the dispersion relation of the fundamental mode of BSCCO PSW is shown in Figure 4(a). Compared to the light line (dispersion relation of vacuum, shown with red color), it infers that the waveguide supports a bound mode because the waveguide lines are on the right side of the light line [43]. With reducing the temperature, the slope of the curves becomes sharper. Therefore, cooling the waveguide results in reducing the refractive index. Lesser $N_{eff}$ for lower temperature dictates the shorter SPP's wavelength. Moreover, the dispersion curves show that the energy is less confined at lower frequencies due to approaching the light line.

For further investigating the thermal tuning of confinement of the modal energy within the slot, the electric field distribution at 5 nm above the waveguide's surface along the dashed line is shown in the bottom panel of Figure 4 (b), for selected temperatures, at $f$=0.1 THz, and for the optimized waveguide size. Here, the top curve is aligned vertically with respect to the cross-sectional view of the waveguide. The pale blue area in the schematic of the waveguide and within the curve of Figure 4(b) shows the air slot area, while the grey area shows the BSCCO thin-film area. The electric field is enhanced within the slot for all temperatures, whereas it grows as temperature reduces due to the reduction of the loss rate. The electric field distribution at different temperatures in Figure 4(c) also shows the enhancement of field within the slot with reducing the temperature.

The comparison between the BSCCO waveguide and gold waveguide with the same structural size of $h$=300 nm and $w$=100 nm at the frequency of $f$=0.1 THz shows that the BSCCO waveguide has larger energy confinement and also larger propagation length (see Figure S1). Therefore, the propagation characteristics of the proposed BSCCO waveguide are better than the gold waveguide. Above $T_c$, the propagation length of SPPs for the BSCCO waveguide is comparable to those of the gold plasmonic waveguide (See Figure S1 in the supplementary material).

**Conclusions**

We numerically investigated the temperature-dependent modal characteristics of high-$T_c$ superconducting BSCCO plasmonic slot waveguide, including the refractive index, propagation length, and the mode confinement in the slot region at the mm-waves to THz range of frequency. We showed that the propagation length of SPP increases as material enters the superconducting phase. In addition, we investigated the frequency tuning of the modal characteristics. Compared with the gold waveguide, the BSCCO waveguide at $T$=10 K offers higher mode confinement within the gap and a larger propagation length below $T_c$. The proposed BSCCO plasmonic waveguide helps realize a fully integrated BSCCO THz circuitry for applications in cryogenic on-chip quantum communication and low-loss data processing.

**References**
1. Delfanazari, K.; Klemm, R.A.; Joyce, H.J.; Ritchie, D.A.; Kadowaki, K. Integrated, Portable, Tunable, and Coherent Terahertz Sources and Sensitive Detectors Based on Layered Superconductors. *Proc. IEEE* **2020**, *108*, 721-734 doi:10.1109/JPROC.2019.2958810.
2. Borodianskyi, E.A.; Krasnov, V.M. Josephson emission with frequency span 1-11 THz from small $Bi_2Sr_2CaCu_2O_{8+\delta}$ mesa structures. *Nat. Commun.* **2017**, *8*, 1–7, doi:10.1038/s41467-017-01888-4.
3. Ozyuzer, L.; Koshelev, A.E.; Kurter, C.; Gopalsami, N.; Li, Q.; Tachiki, M.; Kadowaki, K.; Yamamoto, T.; Minami, H.; Yamaguchi, H.; et al. Emission of coherent THz radiation from superconductors. *Science.* **2007**, *318*, 1291–1293, doi:10.1126/science.1149802.
4. Ozyuzer, L.; Simsek, Y.; Koseoglu, H.; Turkoglu, F.; Kurter, C.; Welp, U.; Koshelev, A.E.; Gray, K.E.; Kwok, W.K.; Yamamoto, T.; et al. Terahertz wave emission from intrinsic Josephson junctions in high-Tc superconductors. *Supercond. Sci. Technol.* **2009**, *22*, 114009 doi:10.1088/0953-2048/22/11/114009.
5. Welp, U.; Kadowaki, K.; Kleiner, R. Superconducting emitters of THz radiation. *Nat. Photonics* **2013**, *7*, 702–710, doi:10.1038/nphoton.2013.216.
6. Tsujimoto, M.; Yamamoto, T.; Delfanazari, K.; Nakayama, R.; Kitamura, T.; Sawamura, M.; Kashiwagi, T.; Minami, H.; Tachiki, M.; Kadowaki, K.; et al. Broadly tunable subterahertz emission from Internal Branches of the current-Voltage characteristics of superconducting $Bi_2Sr_2CaCu_2O_{8+\delta}$ single crystals. *Phys. Rev. Lett.* **2012**, *108*, 107006, doi:10.1103/PhysRevLett.108.107006.
7. Klemm, R.A.; Delfanazari, K.; Tsujimoto, M.; Kashiwagi, T.; Kitamura, T.; Yamamoto, T.; Sawamura, M.; Ishida, K.; Hattori, T.; Kadowaki, K. Modeling the electromagnetic cavity mode contributions to the THz emission from triangular $Bi_2Sr_2CaCu_2O_{8+\delta}$ mesas. *Phys. C* **2013**, *491*, 30–34. doi: 10.1016/j.physc.2012.11.006.
8. Delfanazari, K.; Asai, H.; Tsujimoto, M.; Kashiwagi, T.; Kitamura, T.; Yamamoto, T.; Sawamura, M.; Ishida, K.; Tachiki, M.; Klemm, R.A.; et al. Study of coherent and continuous terahertz wave emission in equilateral triangular mesas of superconducting $Bi_2Sr_2CaCu_2O_{8+\delta}$ intrinsic Josephson junctions. *Phys. C Supercond. its Appl.* **2013**, *491*, 16–19, doi:10.1016/j.physc.2012.12.009.
9. Kitamura, T.; Kashiwagi, T.; Tsujimoto, M.; Delfanazari, K.; Sawamura, M.; Ishida, K.; Sekimoto, S.; Watanabe, C.; Yamamoto, T.; Minami, H.; et al. Effects of magnetic fields on the coherent THz emission from mesas of single crystal $Bi_2Sr_2CaCu_2O_{8+\delta}$. *Phys. C Supercond. its Appl.* **2013**, *494*, 117–120, doi:10.1016/j.physc.2013.05.011.
10. Delfanazari, K.; Asai, H.; Tsujimoto, M.; Kashiwagi, T.; Kitamura, T.; Yamamoto, T.; Sawamura, M.; Ishida, K.; Watanabe, C.; Sekimoto, S.; et al. Tunable terahertz emission from the intrinsic Josephson junctions in acute isosceles triangular $Bi_2Sr_2CaCu_2O_{8+\delta}$ mesas. *Opt. Express* **2013**, *21*, 2171-2184, doi:10.1364/oe.21.002171.
11. Kadowaki, K.; Tsujimoto, M.; Delfanazari, K.; Kitamura, T.; Sawamura, M.; Asai, H.; Yamamoto, T.; Ishida, K.; Watanabe, C.; Sekimoto, S.; et al. Quantum terahertz electronics (QTE) using coherent radiation from high temperature superconducting $Bi_2Sr_2CaCu_2O_{8+\delta}$ intrinsic Josephson junctions. *Phys. C Supercond. its Appl.* **2013**, *491*, 2–6, doi:10.1016/j.physc.2013.04.011.
12. Delfanazari, K.; Asai, H.; Tsujimoto, M.; Kashiwagi, T.; Kitamura, T.; Ishida, K.; Watanabe, C.; Sekimoto, S.; Yamamoto, T.; Minami, H.; et al. Terahertz oscillating devices based upon the intrinsic Josephson junctions in a high temperature superconductor. *J. Infrared, Millimeter, Terahertz Waves* **2014**, *35*, 131–146, doi:10.1007/s10762-013-0027-y.
13. Delfanazari, K.; Asai, H.; Tsujimoto, M.; Kashiwagi, T.; Kitamura, T.; Yamamoto, T.; Wilson, W.; Klemm, R.A.; Hattori, T.; Kadowaki, K. Effect of bias electrode position on terahertz radiation from pentagonal mesas of superconducting $Bi_2Sr_2CaCu_2O_{8+\delta}$. *IEEE Trans. Terahertz Sci. Technol.* **2015**, *5*, 505–511, doi:10.1109/TTHZ.2015.2409552.

14. Cerkoney, D.P.; Reid, C.; Doty, C.M.; Gramajo, A.; Campbell, T.D.; Morales, M.A.; Delfanazari, K.; Tsujimoto, M.; Kashiwagi, T.; Yamamoto, T.; et al. Cavity mode enhancement of terahertz emission from equilateral triangular microstrip antennas of the high-T c superconductor $Bi_2Sr_2CaCu_2O_{8+\delta}$. *J. Phys. Condens. Matter* **2017**, *29*, 015601, doi:10.1088/0953-8984/29/1/015601.
15. Kashiwagi, T.; Yamamoto, T.; Minami, H.; Tsujimoto, M.; Yoshizaki, R.; Delfanazari, K.; Kitamura, T.; Watanabe, C.; Nakade, K.; Yasui, T.; et al. Efficient Fabrication of Intrinsic-Josephson-Junction Terahertz Oscillators with Greatly Reduced Self-Heating Effects. *Phys. Rev. Appl.* **2015**, *4*, 054018, doi:10.1103/PhysRevApplied.4.054018.
16. Kashiwagi, T.; Yamamoto, T.; Kitamura, T.; Asanuma, K.; Watanabe, C.; Nakade, K.; Yasui, T.; Saiwai, Y.; Shibano, Y.; Kubo, H.; et al. Generation of electromagnetic waves from 0.3 to 1.6 terahertz with a high- Tc superconducting $Bi_2Sr_2CaCu_2O_{8+\delta}$ intrinsic Josephson junction emitter. *Appl. Phys. Lett.* **2015**, *106*, 092601, doi:10.1063/1.4914083.
17. Delfanazari, K.; Tsujimoto, M.; Kashiwagi, T.; Yamamoto, T.; Nakayama, R.; Hagino, S.; Kitamura, T.; Sawamura, M.; Hattori, T.; Minami, H.; et al. THz emission from a triangular mesa structure of Bi-2212 intrinsic Josephson junctions. *J. Phys. Conf. Ser.* **2012**, *400*, 022014, doi:10.1088/1742-6596/400/2/022014.
18. Delfanazari, K.; Asai, H.; Tsujimoto, M.; Kashiwagi, T.; Kitamura, T.; Sawamura, M.; Ishida, K.; Yamamoto, T.; Hattori, T.; Klemm, R.A.; et al. Experimental and theoretical studies of mesas of several geometries for terahertz wave radiation from the intrinsic Josephson junctions in superconducting $Bi_2Sr_2CaCu_2O_{8+\delta}$. *37th Int. Conf. Infrared, Millimeter, Terahertz Waves, IRMMW-THz* **2012**, 1–2, doi:10.1109/IRMMW-THz.2012.6380230.
19. Kashiwagi, T.; Tsujimoto, M.; Yamamoto, T.; Minami, H.; Yamaki, K.; Delfanazari, K.; Deguchi, K.; Orita, N.; Koike, T.; Nakayama, R.; et al. High temperature superconductor terahertz emitters: Fundamental physics and its applications. *Jpn. J. Appl. Phys.* **2012**, *51*, 010113, doi:10.1143/JJAP.51.010113.
20. Savinov, V.; Delfanazari, K.; Fedotov, V.A.; Zheludev, N.I. Giant sub-THz nonlinear response in superconducting metamaterial. *Conf. Lasers Electro-Optics Eur. - Tech. Dig.* **2014**, *SW3I-8*, 2–3, doi:10.1364/cleo_si.2014.sw3i.8.
21. Klemm, R.A.; Davis, A.E.; Wang, Q.X.; Yamamoto, T.; Cerkoney, D.P.; Reid, C.; Koopman, M.L.; Minami, H.; Kashiwagi, T.; Rain, J.R.; et al. Terahertz emission from the intrinsic Josephson junctions of high-symmetry thermally-managed Bi2Sr2CaCu2O8+δ microstrip antennas. *IOP Conf. Ser. Mater. Sci. Eng.* **2017**, *279*, doi:10.1088/1757-899X/279/1/012017.
22. Tsujimoto, M.; Minami, H.; Delfanazari, K.; Sawamura, M.; Nakayama, R.; Kitamura, T.; Yamamoto, T.; Kashiwagi, T.; Hattori, T.; Kadowaki, K. Terahertz imaging system using high-Tc superconducting oscillation devices. *J. Appl. Phys.* **2012**, *111*, 123111, doi:10.1063/1.4729799.
23. Delfanazari, K.; Klemm, R.A.; Tsujimoto, M.; Cerkoney, D.P.; Yamamoto, T.; Kashiwagi, T.; Kadowaki, K. Cavity modes in broadly tunable superconducting coherent terahertz sources. *J. Phys. Conf. Ser.* **2019**, *1182*, 012011, doi:10.1088/1742-6596/1182/1/012011.
24. T Kashiwagi, K Deguchi, M Tsujimoto, T Koike, N Orita, K Delfanazari, R Nakayama, T Kitamura, S Hagino, M Sawamura, T Yamamoto, H.M. and K.K. Excitation mode characteristics in Bi2212 rectangular mesa structures. *J. Phys. Conf. Ser. 400* **2012**, *400*, 022050, doi:10.1088/1742-6596/400/2/022050.
25. M Tsujimoto, T Yamamoto, K Delfanazari, R Nakayama, N Orita, T Koike, K Deguchi, T Kashiwagi, H.M. and K.K. THz-wave emission from inner I-V branches of intrinsic Josephson junctions in $Bi_2Sr_2CaCu_2O_{8+\delta}$. *J. Phys. Conf. Ser.* **2012**, *400*, 022127, doi:10.1088/1742-6596/400/2/022127.
26. Rahmonov, I.R.; Shukrinov, Y.M.; Zemlyanaya, E. V.; Sarhadov, I.; Andreeva, O. Mathematical modeling of intrinsic Josephson junctions with capacitive and inductive couplings. *J. Phys. Conf. Ser.* **2012**, *393*, 012022, doi:10.1088/1742-6596/393/1/012022.
27. Botha, A.E.; Rahmonov, I.R.; Shukrinov, Y.M. Spontaneous and Controlled Chaos Synchronization in Intrinsic Josephson Junctions. *IEEE Trans. Appl. Supercond.* **2018**, *28*, 1-6, doi:10.1109/TASC.2018.2841902.
28. Nakade, K.; Kashiwagi, T.; Saiwai, Y.; Minami, H.; Yamamoto, T.; Klemm, R.A.; Kadowaki, K. Applications using high-tc superconducting terahertz emitters. *Sci. Rep.* **2016**, *6*, 1–8, doi:10.1038/srep23178.
29. Singh, R.; Zheludev, N. Superconductor photonics. *Nat. Photonics* **2014**, *8*, 679–680, doi:10.1038/nphoton.2014.206.
30. Hayashi, S.; Okamoto, T. Plasmonics: Visit the past to know the future. *J. Phys. D. Appl. Phys.* **2012**, *45*, 433001 doi:10.1088/0022-3727/45/43/433001.
31. Barnes, W.L.; Dereux, A.; Ebbesen, T.W. Surface plasmon subwavelength optics. *Nature* **2003**, *424*, 824–830, doi:10.1038/nature01937.
32. Tsiatmas, A.; Fedotov, V.A.; García De Abajo, F.J.; Zheludev, N.I. Low-loss terahertz superconducting plasmonics. *New J. Phys.* **2012**, *14*, 115006, doi:10.1088/1367-2630/14/11/115006.
33. Ma, Y.; Eldlio, M.; Maeda, H.; Zhou, J.; Cada, M.; Zhai, X.; Wang, L.L.; Wang, L.L.; Lindquist, N.C.; Nagpal, P.; et al. Plasmonic properties of superconductor–insulator–superconductor waveguide. *Appl. Phys. Express* **2016**, *9*, 072201. doi: 10.7567/APEX.9.072201.
34. Veronis, G.; Fan, S. Modes of subwavelength plasmonic slot waveguides. *J. Light. Technol.* **2007**, *25*, 2511–2521, doi:10.1109/JLT.2007.903544.
35. **Novotny, L. and Hecht, B**, *Principles of nano-optics* Cambridge university press, 2012; ISBN 9781107005464.
36. Kalhor, S.; Ghanaatshoar, M.; Delfanazari, K. Guiding of terahertz photons in superconducting nano-circuits. *2020 Int. Conf. UK-China Emerg. Technol. UCET 2020* **2020**, 1-3, doi:10.1109/UCET51115.2020.9205480.
37. Corson, J.; Orenstein, J.; Oh, S.; O'Donnell, J.; Eckstein, J.N. Nodal quasiparticle lifetime in the superconducting state of $Bi_2Sr_2CaCu_2O_{8+\delta}$. *Phys. Rev. Lett.* **2000**, *85*, 2569–2572, doi:10.1103/PhysRevLett.85.2569.
38. Mallozzi, R.; Corson, J.; Orenstein, J.; Eckstein, J.N.; Bozovic, I. Terahertz conductivity and c-axis plasma resonance in $Bi_2Sr_2CaCu_2O_{8+\delta}$. *J. Phys. Chem. Solids* **1998**, *59*, 2095–2099, doi:10.1016/S0022-3697(98)00181-4.

39. Kalhor, S.; Ghanaatshoar, M.; Kashiwagi, T.; Kadowaki, K.; Kelly, M.J.; Delfanazari, K. Thermal Tuning of High-Tc Superconducting Bi2Sr2CaCu2O8+δ Terahertz Metamaterial. *IEEE Photonics J.* **2017**, *9*, 1–8, doi:10.1109/JPHOT.2017.2754465.
40. Barnes, W.L. Surface plasmon-polariton length scales: A route to sub-wavelength optics. *J. Opt. A Pure Appl. Opt.* **2006**, *8*, S87–S93, doi:10.1088/1464-4258/8/4/S06.
41. Tian, Z.; Singh, R.; Han, J.; Gu, J.; Xing, Q.; Zhang, W. Terahertz superconducting plasmonic hole array. *Opt. Lett.* **2010**, *35*, 3586–3588, doi:10.1364/OL.35.003586.
42. Tsiatmas, A.; Buckingham, A.R.; Fedotov, V.A.; Wang, S.; Chen, Y.; De Groot, P.A.J.; Zheludev, N.I. Superconducting plasmonics and extraordinary transmission. *Appl. Phys. Lett.* **2010**, *97*, 111106, doi:10.1063/1.3489091.
43. MAIER, S.A. *Plasmonics fundamentals and applications*; Springer Science & Business Media., 2007; ISBN 9780387331508.

**Supplementary Information:**

The conductivity of gold ($\sigma_{Au}$) is described by the Drude model expression as

$$\sigma_{Au} = \varepsilon_0 \frac{\omega_p^2}{\gamma + i\omega} \qquad (S1)$$

where plasma frequency $\omega_p$ is $2\pi \times 2175$ THz and collision frequency $\gamma$ is $2\pi \times 6.5$ THz [1]. Here, $\varepsilon_0$ is the vacuum electric constant. It should be noted that no temperature tuning is expected for the gold waveguide.

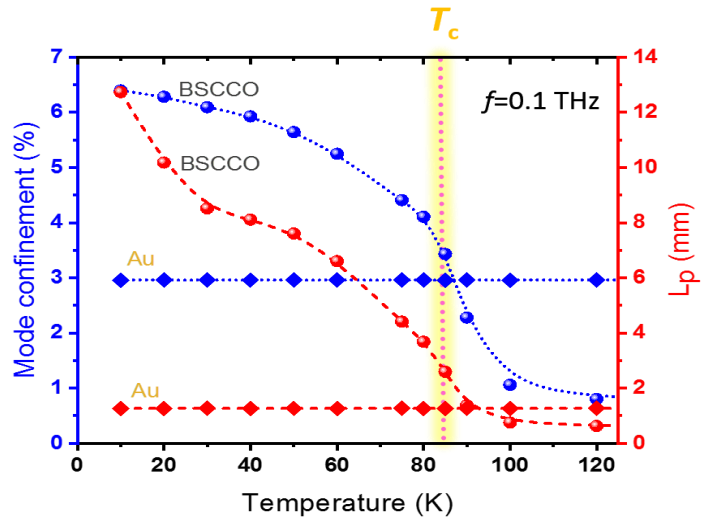

**Figure S1**: Mode confinement and propagation length of SPPs ($L_p$) of BSCCO and gold plasmonic waveguide with structural size of *h*=300 n and *w*=300 nm at *f*=0.1 THz. The vertical dashed line shows the transition temperature of BSCCO. Modal properties of gold waveguide in not dependent on temperature.

[1] Ordal, M.A.; Long, L.L.; Bell, R.J.; Bell, S.E.; Bell, R.R.; Alexander, R.W.; Ward, C.A. Optical properties of the metals Al, Co, Cu, Au, Fe, Pb, Ni, Pd, Pt, Ag, Ti, and W in the infrared and far infrared. *Appl. Opt.* **1983**, *22*, 1099-1119, doi:10.1364/AO.22.001099.